# The False Positive Risk: a proposal concerning what to do about *p*-values

David Colquhoun

University College London, Gower Street, London WC1E 6BT

d.colquhoun@ucl.ac.uk

## Contents

## Abstract

It is widely acknowledged that the biomedical literature suffers from a surfeit of false positive results.  Part of the reason for this is the persistence of the myth that observation of p < 0.05 is sufficient justification to claim that you've made a discovery.

It is hopeless to expect users to change their reliance on *p*-values unless they are offered an alternative way of judging the reliability of their conclusions.  If the alternative method is to have a chance of being adopted widely, it will have to be easy to understand and to calculate. One such proposal is based on calculation of false positive risk.

It is suggested that *p* values and confidence intervals should continue to be given, but that they should be supplemented by a single additional number that conveys the strength of the evidence better than the *p* value.  This number could be the minimum false positive risk (that calculated on the assumption of a prior probability of 0.5, the largest value that can be assumed in the absence of hard prior data).  Alternatively one could specify the prior probability that it would be necessary to believe in order to achieve a false positive risk of, say, 0.05.

## 1. Introduction

It is a truth universally acknowledged that some areas of science suffer from a surfeit of false positives.

By a false positive, I mean that you claim that an effect exists when in fact the results could easily have occurred by chance only.

Every false positive that gets published in the peer-reviewed literature provides ammunition for those who oppose science. That has never been more important. There is now strong pressure from within science as well as from those outside it, to improve reliability of the published results.

In the end, the only way to solve the problem of reproducibility is to do more replication and to reduce the incentives that are imposed on scientists to produce unreliable work.  The publish-or-perish culture has damaged science, as has the judgement of their work by silly metrics.

In the meantime, improvements in statistical practice can help.

In this paper I am referring to the practice of experimentalists, and to the statistical advice given by journal editors, when they do not have a professional statistician as part of the team. The vast majority of published tests of significance are in this category.

The aim is to answer the following .question; if you observe a 'significant' *p*-value after doing a single unbiased experiment, what is the probability that your result is a false positive? In this paper, we expand on the proposals made by Colquhoun (2017, 2018), and compare them with some other methods that have been proposed.

### 2. How can statistical practice by users and journals be changed?

In the biomedical literature it is still almost universal to do a test of statistical significance, and if the resulting *p* value is less than 0.05 to declare that the result is "statistically significant" and to conclude that the claimed effect is likely to be real, rather than chance. It has been known for many decades that this procedure is unsafe. It can lead to many false positives.

Many users still think that the *p* value is the probability that your results are due to chance (Gigerenzer *et al.* 2004). That, it seems, is what they want to know. But that isn't what the *p*-value tells you.

There can be no doubt that the teaching of statistics to users has failed to improve this practice.

> "The standard approach in teaching, of stressing the formal definition of a *p*-value while warning against its misinterpretation, has simply been an abysmal failure" (Sellke, Bayarri and Berger 2001, p71)

The ASA statement (Wasserstein and Lazar 2016) gave a cogent critique of *p*-values. But it failed to come to any conclusions about what should be done. That is why it's unlikely to have much effect on users of statistics. The reason for the lack of recommendations was the inability of statisticians to agree. In the opening words of Kristin Lennox, at a talk she gave at Lawrence Livermore labs,

> "Today I speak to you of war. A war that has pitted statistician against statistician for nearly 100 years. A mathematical conflict that has recently come to the attention of the 'normal' people". Lennox (2016)

I suggest that the only hope of having the much-needed improvement in statistical practice is to adopt some sort of compromise between frequentist and Bayesian approaches. A good candidate for this is to convert the observed *p* value to the false positive risk. This should appeal to users because the false positive risk is what most users still think that *p* values give them. It's the probability that your result occurred by chance only, a very familiar concept.

There is nothing very new about "recalibrating" *p* values, such that they mean what users mistakenly thought they mean. Various ways of doing this have been proposed, notably by Berger & Sellke (1987), Sellke *et al.*(2001), Goodman (1993, 1999 a, b) and by Johnson (2013a,b).

The misinterpretation of *p*-values is, of course, only one of the widespread misunderstandings of statistics. It's relatively easy to teach users about the hazards of multiple comparisons and the problem of *p*-hacking. It seems to be much harder to teach what a *p* value tells you and what it doesn't tell you. The abuse of $p < 0.05$ is still almost universal, at least in the biomedical literature.

Journals still give ghastly advice like

> "a level of probability (*p*) deemed to constitute the threshold for statistical significance should be defined in Methods, and not varied later in Results (by presentation of multiple levels of significance). Thus, ordinarily $p < 0.05$ should be used throughout a paper to denote statistically significant differences between groups." (Curtis et al. 2014)

Many people have advised culture change. Desirable though that is, such pleading has had very little effect. Others have said that authors should ask a professional statistician. That is obviously desirable too, but there are nowhere near enough statisticians to make that possible, and especially not enough statisticians who are willing to spend time analysing your data.

The constraints on improvement are as follows.

1. Users will continue to use statistical tests, with the aim of trying to avoid making fools of themselves by publishing false positive results. That's entirely laudable.

2. Most users won't have access to professional statistical advice. There just isn't enough of it (in a form that users can understand, and put into practice).

3. The only way to stop people claiming a discovery whenever they observe $p < 0.05$ is to give them an alternative procedure.

4. Full Bayesian analysis will probably never be adopted as a routine substitute for *p* values. It has a place in formal clinical trials that are guided by a professional statistician, but it's too complicated for most users, and experimentalists distrust (rightly, in my view) informative priors. As Valen Johnson said

> "… subjective Bayesian testing procedures have not been - and will likely never be - generally accepted by the scientific community" (Johnson 2013b)

A special case of Bayesian analysis, the *false positive risk approach*, is simple enough to be widely understood and allows the *p* value to be supplemented by a single number that gives a much better idea of the strength of the evidence than a *p* value alone.

### 3. The false positive risk approach

I'll consider here only the question of how to avoid false positives when interpreting the result of a single unbiased test of significance. All the other problems of multiple comparisons, inadequate randomisation or blinding, *p*-hacking etc. can only make matters still worse in real life.

To proceed, we need to define more precisely "the probability that your results are due to chance". This should be defined as the probability, in the light of the *p* value that you observe, you declare that an effect is real, when in fact, it isn't. It's the risk of getting a false positive. This quantity has been given various names. For example it's been called the false positive report rate (FPRP) by Wacholder *et al.*(2004) though they used the *p-less-than* approach (see Colquhoun, 2017, section 3) which is inappropriate for interpretation of single tests. It was called the false discovery rate by Colquhoun (2014, 2016a, 2016b, 2018) –this was a foolish choice because it risks confusion with the use of that term in the context of multiple comparison problems (e.g. Benjamini & Hochberg, 1995). A better term is the false positive rate, or, better still, false positive risk (FPR). The latter term was recommended by Prof D. Spiegelhalter (personal communication), because we are talking about interpretation of a single experiment: we are not trying to estimate long term error rates.

It should be noted that corrections for multiple comparisons aim to correct only the type 1 error rate. The result is essentially a corrected *p* value and is therefore open to misinterpretation in the same way as any other *p* value.

In order to calculate the false positive risk (FPR) it's necessary to invoke Bayes' theorem. This can be expressed in terms of odds as

*posterior odds on $H_1$ = Bayes factor X prior odds*     (1)

Or, in symbols,

$$\frac{P(H_1|\,data)}{P(H_0|\,data)} = \frac{P(data|H_1)}{P(data|H_0)} \times \frac{P(H_1)}{P(H_0)}$$

(2)

where $H_0$ is the null hypothesis and $H_1$ is the alternative hypothesis.

The bad news is that there are many ways to calculate the Bayes factor (see e.g. Held & Ott, 2018). The good news is that many of them give similar results, within a factor of 2 or so, so whichever method is used, the conclusions are not likely to differ much in practice. In order to simulate *t* tests (e.g as in Colquhoun 2014) it is necessary to specify an alternative hypothesis, e.g. that the true effect size is 1 standard deviation. As we shall see below (section 5, Figure 1), this does not reduce the generality of the conclusions as much as it might appear at first sight. Thus, we choose to test

$$H_0: \theta = 0 \text{ against } H_1: \theta = \theta_1 \neq 0$$

One advantage of using a simple alternative hypothesis is that the Bayes factor becomes a simple likelihood ratio, and there is no need to postulate a prior *distribution*, but rather a simple prior probability suffices: $P(H_1) = 1 - P(H_0)$ is the probability that there is a real (non-zero) effect size. Thus (2) can be written as

$$\frac{P(H_1|\,data)}{1 - P(H_1|\,data)} = L_{10} \times \frac{P(H_1)}{1 - P(H_1)}$$

(3)

where the likelihood ratio in favour of $H_1$, relative to $H_0$, is defined as $LR_{10}$ (following the notation of Held & Ott, 2018).

$$L_{10} = \frac{P(data|H_1)}{P(data|H_0)}$$

(4)

It has often been argued that the likelihood ratio alone is a better way than a *p* value to represent the strength of evidence against the null hypothesis, e.g. Goodman (1993, 1999a,b).

It would indeed be a big advance if likelihood ratios were cited, instead of, or as well as, *p* values. One problem with this proposal is that very few non-professional users of statistics are familiar with the idea of likelihood ratios as evidence. Another problem is that it's hard to judge how big the likelihood ratio must be before it's safe to claim that you have good evidence against the null hypothesis. It seems better to specify the result as a false positive risk, if only because most users still think, mistakenly, that that is what the *p*-value tells them (e.g. Gigerenzer *et al.*, 2004). The idea is already familiar to users. That gives it a chance of being adopted. The false positive risk (FPR) is

$$P(H_0|\,data) = 1 - P(H_1|\,data)$$

so, from (3) and (4) we get

$$FPR = \cfrac{1}{1 + L_{10} \cfrac{P(H_1)}{1 - P(H_1)}}$$

(5)

Thus, in the case where the prior probability, $P(H_1) = 0.5$, the false positive risk is seen to be just another way of expressing the likelihood ratio:

$$FPR = \frac{1}{1 + L_{10}}$$

Notice that we have chosen to write the likelihood ratio as the odds on $H_1$. The bigger the value, the more probable is the data given $H_1$, relative to $H_0$. Often the likelihood ratio is written the other way up, as $LR_{01} = 1/LR_{10}$, so large values favour $H_0$. References to minimum values of $LR_{01}$ –the smallest evidence for $H_0$ –are therefore equivalent to maximum values for $LR_{10}$, the greatest evidence for there being a real non-zero effect that is provided by the data,

In the case of a simple alternative hypothesis, the likelihood ratio can be calculated in two different ways: the *p-equals* approach or the *p-less-than* approach (e.g. Goodman, 1993, 1999b, and section 3 in Colquhoun, 2017). Once the experiment is done, the *p* value that it produces is part of the observations. We are interested only in the particular *p* value that was observed, not in smaller *p* values that were not observed. Since the aim of this work is to interpret a single *p* value from a single experiment (rather than trying to estimate the long term error rate), the *p-equals* method is the appropriate way to calculate likelihood ratios.

For small samples, for which a *t* test is appropriate, the likelihood ratio can be calculated as a ratio of probability densities (see Appendix). The effect of sample size on false positive risk was investigated by Colquhoun (2017).

Bayes' theorem requires that we specify a prior probability and in the vast majority of cases, we don't know what it is. But I don't think that that means we have to give up on the FPR.

There are two ways to circumvent the inconvenient fact that we can't give a value for the prior probability that there's a real effect. One is to assume that the prior probability of a real effect, *P*(H1), is 0.5, i.e. that there is a 50:50 chance of there being a real effect before the experiment was done. The other way is to calculate the prior that you would need to believe in order to achieve an FPR of, say 5%. This so-called reverse Bayes approach was mentioned by Good (1950) and has been used, for example, by Carlin & Louis (1996), Matthews (2001, 2018), Held (2013) and Colquhoun (2017, 2018). In both cases it would be up to you to persuade your readers and journal editors that the value of the prior probability is reasonable. Inductive inference inevitably has an element of subjectivity (Colquhoun 2016).

The proposal to specify the FPR is likely to be more acceptable to users than most because it uses concepts which are familiar to most users. It starts by calculating *p*-values and uses the familiar point null hypothesis. The idea of prior probability can be explained via diagnostic screening tests and the results can be explained to users by simulation of tests (e.g. Colquhoun 2014).

Exact calculations can be done using R programs (Colquhoun 2017), or with a web calculator ([Colquhoun and Longstaff 2017](#)). The variables involved are the observed *p* value, the false positive risk (FPR) and the prior probability that there is a real effect. If any two of these are specified, the third can be calculated. The calculations give the same FPR if the number of observations and the normalised effect size are adjusted to keep the power constant (Figure 1 and Colquhoun 2017), so when doing the calculations these should be chosen to match the power of your experiment: see section 5, below.

### 4. The point null

Some people don't like the assumption of a point null that's made in this proposed approach to calculating the false positive risk, but it seems natural to users who wish to know how likely their observations would be if there were really no effect (i.e. if the point null were true). It's often claimed that the null is never exactly true. This isn't necessarily so (just give identical treatments to both groups). But more importantly, it doesn't matter. We aren't saying that the effect size is exactly zero: that would obviously be impossible. We are looking at the likelihood ratio, i.e. at the probability of the data if the true effect size were not zero relative to the probability of the data if the effect size were zero. If the latter probability is bigger than the former then clearly we can't be sure that there is anything but chance at work. This does *not* mean that we are saying that the effect size is exactly zero. From the point of view of the experimenter, testing the point null makes total sense. In any case, it makes little difference if the null hypothesis is a narrow band centred on zero (Berger and Delampady 1987).

### 5. The simple alternative hypothesis

It was stated above that the need to specify a fixed alternative hypothesis is not as restrictive as it seems at first. This is the case because false positive risk is essentially independent of the effect size if the power of the experiment is constant. This is illustrated in Figure 1, which shows the result of plotting FPR against the true normalised effect size. In this graph the power is kept constant by changing the sample size, *n*, for each point. The power is calculated in the standard way, for the normalised effect size and $p = 0.05$. When the FPR is calculated by the *p-less-than* method it is exactly independent of the effect size. When the FPR is calculated by the *p-equals* method, as is appropriate for our problem, the FPR is essentially independent of effect size up to an effect size of about 1 (standard deviation) and declines only slightly for larger effects. This is the basis for the recommendation that the inputs for the calculator should match the power of the actual experiment.

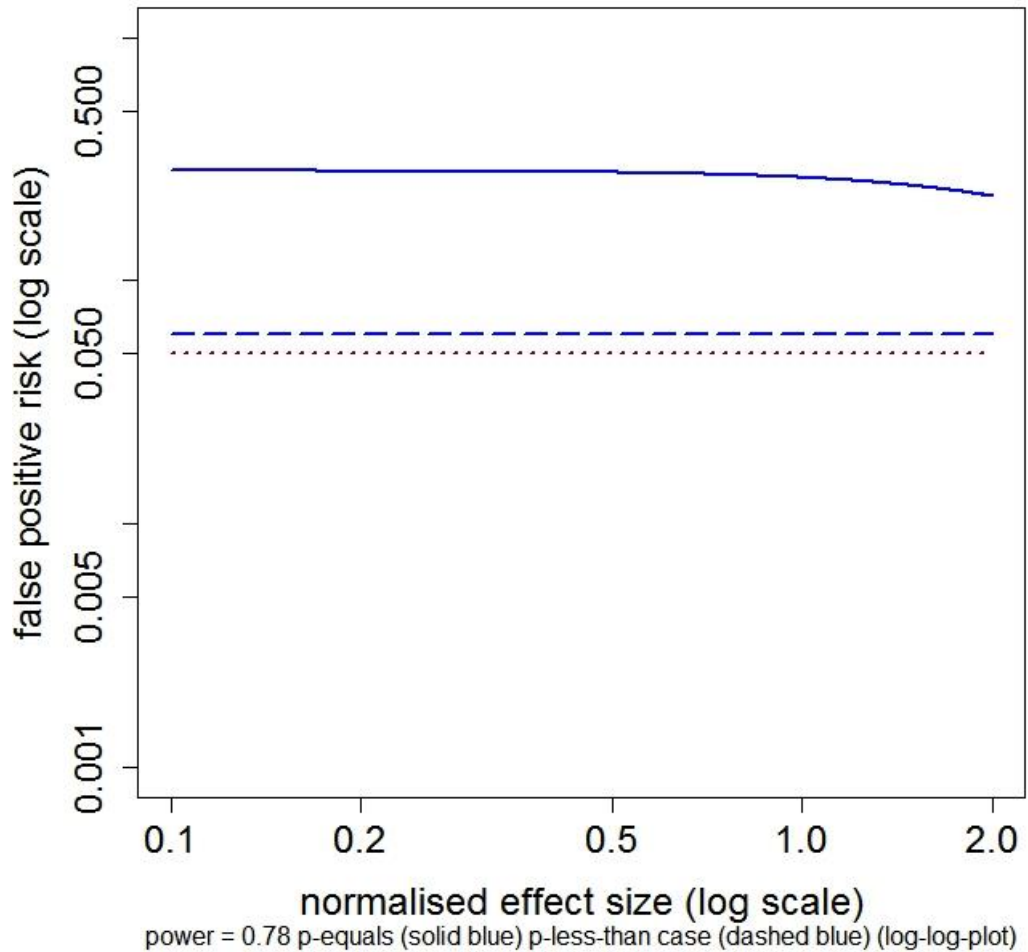

Figure 1

Plot of false positive risk against normalised effect size, with power kept constant throughout the curves by varying *n*. The dashed blue line shows the FPR calculated by the *p-less-than* method. The solid blue line shows the FPR calculated by the *p-equals* method. This example is calculated for an observed *p* value of 0.05, with power kept constant at 0.78 (power calculated conventionally at *p* = 0.05), with prior probability $P(H_1)$ =0.5. The sample size needed to keep the power constant at 0.78 varies from *n* = 1495 at effect size =0.1, to *n* = 5 at effect size = 2.0 (the range of plotted values). The dotted red line marks an FPR of 0.05, the same as the observed *p* value. Calculated with *Plot-FPR-v-ES-constant-power.R*, output file: *FPR-vs-ES-const-power.txt* (supplementary material).

## 6. The Jeffreys – Lindley 'paradox'

It may be surprising, at first sight, that the FPR approaches 100 percent as the sample size is increased when it is calculated by the *p-equals* method, whereas the FPR decreases monotonically with sample size when it's calculated by the *p-less-than* method. The former case is illustrated in Figure 2. As sample size increases, the evidence in favour of the null hypothesis declines at first but then increases.

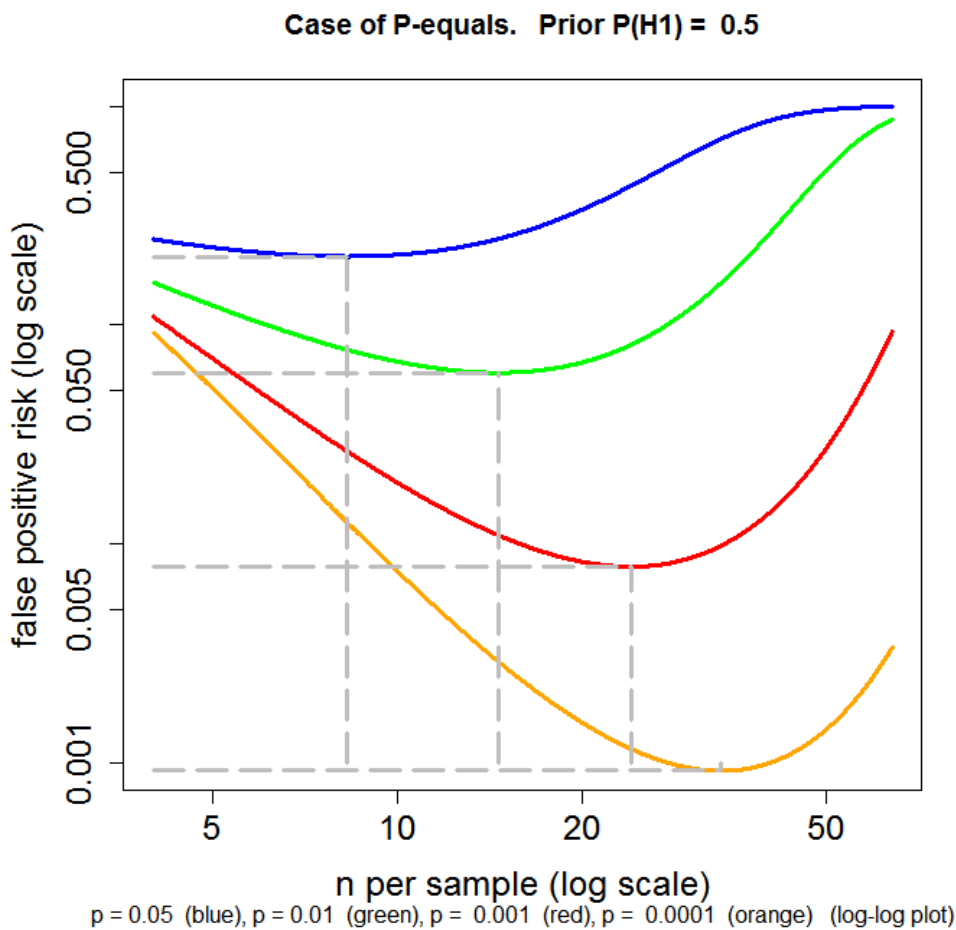

Figure 2

False positive risk plotted against *n*, the number of observations per group for a two independent sample *t* test, with normalised true effect size of 1 standard deviation. . The false positive risk is calculated by the *p-equals* method, with a prior probability $P(H_1) = 0.5$ (Appendix eq. A6). Log-log plot. Calculations for four different observed *p* values, from top to bottom these are: $p = 0.05$ (blue), $p = 0.01$ (green), $p = 0.001$ (red), $p = 0.0001$ (orange). The power of the tests varies throughout the curves. For example, for the $p = 0.05$ curve, the power is 0.22 for $n = 4$ and power is 0.9999 for $n = 64$ (the extremes of the plotted range). The minimum FPR (marked with grey dashed lines) is 0.206 at $n = 8$. This may be compared with the FPR of 0.27 at $n = 16$, at which point the power is 0.78 (as in Colquhoun, 2014, 2017). Values for other curves are given in the print file for these plots (Calculated with *Plot-FPR-vs-n.R*. Output file: *Plot-FPR-vs-n.txt*. supplementary material)

This phenomenon is an example of the Jeffreys-Lindley 'paradox' (Lindley, 1957, Bartlett, 1957).

It can be explained intuitively as follows.

It's convenient to use the web calculator at http://fpr-calc.ucl.ac.uk/ The third radio button shows that for $p = 0.05$, and normalised effect size of 1, a sample of 16 observations in each group gives a power of 0.78. That is usually considered to be a well-powered experiment. Many published experiments have much lower power. For an observed *p* value of 0.05, the FPR for $n = 16$ and a prior of 0.5 is 27% when calculated by the *p-equals* method. This is one way of looking at the weakness of the evidence against the null hypothesis that's provided by observing $p = 0.05$ in a single well-powered experiment. Another way is to note that the likelihood ratio in favour of there being a real effect (null hypothesis being untrue) is only about 3 to 1 (2.76, see section 5.1 in Colquhoun, 2017). This is much less than the odds of 19:1 that are often wrongly inferred from the observed $p = 0.05$.

Now to explain the apparent paradox. If the sample size is increased from 16 to 64, keeping everything else the same, the FPR increases from 0.78 (at $n = 16$) to 0.996 (at $n = 64$) when calculated by the *p-equals* method .i.e. observation of $p = 0.05$ when *n = 64* provides strong evidence *for* the null hypothesis. But notice that this sample size corresponds to a power of 0.9999, far higher than is ever achieved in practice. The reason for the increased FPR with $n = 64$ can be understood by looking at the corresponding likelihood ratios. If the calculation for $n = 64$ is done by the *p-less-than* method, the likelihood ratio is 20.0 –quite high because most *p*-values will be less than 0.05 in this situation. But the likelihood ratio found by the *p-equals* method has fallen from 2.76 for $n = 16$, to 0.0045 for $n = 64$. In words, the observation (of $p = 0.05$) would be 222 times as likely (1/0.0045) if the null hypothesis were true than it would be if the null hypothesis were false. So observing $p = 0.05$ provides *strong evidence for the null hypothesis* in this case. The reason for this is that with $n = 64$ it's very unlikely that a *p* value *as big as* 0.05 would be observed: 99.99% of *p* values are less than 0.05 (that's the power), and 98.7% of *p* values are less than 0.001 (found using the R script *two_sample-simulation-+LR.R*: supplementary material). With a power of 0.9999, it would be very suspicious indeed to observe a *p*-value *as big as* 0.05.

The resolution of this problem is given in appendix A4 of Colquhoun (2017), and in the Notes tab of the calculator. Make sure that the power matches that of your experiment. If you use a silly power, you'll get a silly answer.

## 7. Some alternative proposals

The suggestion made here is one of several that have been made with the aim of expressing evidence in ways that don't suffer from the same problems as *p* values. They are all based on normal distributions apart from the present proposal (Colquhoun, 2014, 2017) which is based on Student's *t* distribution so that its results are appropriate for the small samples that typify most lab work.

1. Many authors have proposed replacement of null hypothesis testing with full Bayesian analysis. These have their place in, for example, large clinical trials with a professional statistical advisor. But they are unlikely to replace the *p* value in routine use, for two reasons. One is that experimenters will probably never trust the use of informative prior distributions that are not based on hard evidence. The other is that the necessity to test a range of priors makes statement of the results too lengthy and cumbersome for routine use in papers that often cite many *p* values.

2. [Senn (2015)](Senn) has produced an example in which the FPR is close to the *p* value. This uses a smooth prior distribution, rather than the lumped prior used here, and has, so far, been demonstrated only for one-sided tests. This result differs from the results obtained by simulation of repeated *t* tests (Colquhoun, 2014) and is in contrast with the general view that the FPR is bigger, often much bigger, than the *p* value.

3. It has been proposed, as an interim way to reduce the risk of false positives, that the threshold for significance should be reduced from 0.05 to 0.005 (Benjamin *et al*. 2017). This proposal has the huge disadvantage that it perpetuates the dichotomisation into "significant" and "non-significant". And even $p = 0.005$ is not safe for implausible hypotheses (see Colquhoun 2017, section 9, and Table 1). The value of $p = 0.005$ proposed by Benjamin *et al*. (2017) would, in order to achieve a false positive risk of 5%, require a prior probability of real effect of about $P(H_1) = 0.4$ (from the web calculator, with power = 0.78). It is, therefore, safe only for plausible hypotheses. If the prior probability were only $P(H_1) = 0.1$, the false positive risk with $p = 0.005$ would be 24% (from the web calculator: [Colquhoun and Longstaff 2017](Colquhoun)). It would still be unacceptably high even with $p = 0.005$. In order to reduce the FPR to 5% with a prior probability of $P(H_1) = 0.1$, we would need to observe $p = 0.00045$. This conclusion differs from that of Benjamin *et al*. (2017) who say that the $p = 0.005$ threshold, with prior = 0.1, would reduce the false positive risk to 5% (rather than 24%). This is because their calculation is based on the *p-less-than* interpretation which, in my opinion, is not the correct way to look at the problem (see Colquhoun 2017, section 3).

4. Matthews (2018) has proposed a more sophisticated version of the reverse Bayesian proposal made here. He shows that, in the case where there is no objective prior information, the observation of $p = 0.013$ implies 95% credibility, in the Bayesian sense (though Matthews doesn't suggest that this should be used as a revised threshold value). This can be compared with the approach of Colquhoun (2017) which says that when you observe $p$ = 0.013 that you'd need a prior probability of there being a real effect of 0.61 to achieve a false positive risk of 0.05 (with the default inputs, power = 0.78). Or alternatively (using the 3rd radio button on the web calculator), that the FPR is 7.7% for a prior of 0.5.

In practice there isn't much difference between a prior of 0.61 and 0.5 (or between an FPR of 7.7% and 5%), and to that extent both approaches give similar answers. But my approach would give a much bigger FPR for implausible hypotheses (those with priors, $P(H_1)$, less than 0.5). Matthews (2018) would deal with implausible hypotheses in a different way.

5. Bayesian approaches which use a prior distribution that's most likely to reject the null hypothesis. These include Berger and Sellke (1987) and Sellke et al. (2001) – (see Appendix), and V. Johnson's uniformly most powerful Bayesian tests (Johnson 2013a, 2013b), For example, if we observe $p$ = 0.05, Berger *et al.* find a false positive risk of 22 - 29% and Johnson finds 17 – 25%. These are comparable with the FPR of 27% found here for a well-powered study with a prior probability of 0.5.

6. Goodman (1999b) has given an expression for calculating the maximum likelihood ratio in favour of $H_1$ in the case where samples are large enough that the normal distribution, rather than Student's $t$) is appropriate (see Appendix, eq. A9).

Three of these approaches are compared in Figure 3 and in Table 1.

Figure 3 shows the FPR plotted against the observed $p$ value. The solid blue line is calculated as in Colquhoun (2017), for a well-powered experiment ($n$ = 16, normalised effect size = 1 SD), using the *p-equals* method, and prior probability $P(H_1)$ = 0.5 (Appendix eqs. A5 and A6). The dashed blue line shows the FPR calculated by the Sellke-Berger approach (Appendix eqs. A8 and A6), and the dotted blue line is calculated by the Goodman approach (Appendix, eqs.A9 and A6). In all cases the FPR is a great deal bigger than the $p$ value (equality is indicated by the dotted red line). The Goodman approach agrees quite well with that of Colquhoun (2017) for observed $p$ values below 0.05, the range of interest. The Sellke-Berger approach gives larger false positive risks over this range, but doesn't differ enough that the conclusions would differ much in practice.

Table 1 shows values calculated for five different observed $p$ values, shown in column 1. The second column shows the corresponding likelihood ratios, calculated as in Colquhoun (2017), and Appendix eq. A5. These are calculated from the known effect size and standard deviation, but calculations based on the observed effect size and sample standard deviation are similar, especially for $p$ values below 0.01 (see Colquhoun, 2017, section 5.1). For example, the last row of Table 1 shows that if you observed $p$ = 0.001, the likelihood ratio in favour of $H_1$ would be 100, which would normally be regarded as very strong evidence against the null hypothesis. However this is not necessarily true. If the prior probability $P(H_1)$ were only 0.1 then Table 1 shows that the FPR would be 0.08, still bigger than the conventional 5% value. In this case you'd need to observe $p$ = 0.00045 in order to achieve an FPR of 0.05 (e.g. using radio button 2 on the calculator at http://fpr-calc.ucl.ac.uk/ )

The third column of Table 1 shows the result of the reverse Bayesian approach, proposed by Matthews (2001) and by Colquhoun (2017), calculated with Appendix eq, A7. If you observed $p$ = 0.05, then in order to achieve an FPR = 0.05 you would need to persuade your readers that it was reasonable to assume a very high prior probability, $P(H_1)$ = 0.87, that the observed effect was not merely chance. It's unlikely that anyone would believe that in the absence of hard evidence. If you observed $p$ = 0.005 (penultimate row of Table 1) the prior needed to achieve FPR = 0.05 would be 0.4, so you'd be safe if the hypothesis was reasonably plausible.

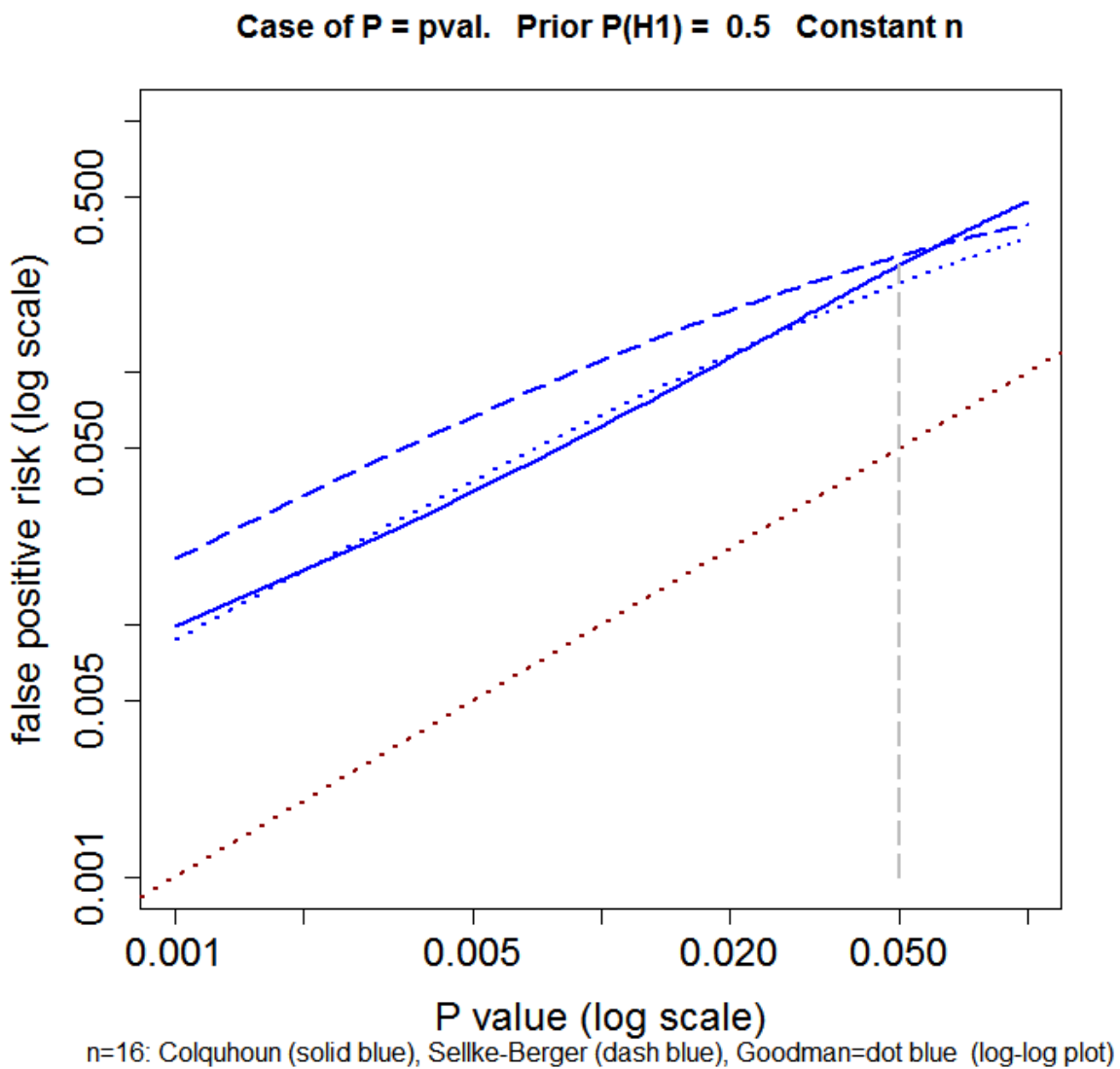

n=16: Colquhoun (solid blue), Sellke-Berger (dash blue), Goodman=dot blue (log-log plot)

Figure 3

Comparison of three approaches to calculation of false positive risk in the case of a simple alternative hypothesis. Solid blue line: the *p-equals* method for *t* tests described in Colquhoun (2017), and appendix eq, A5 and A7. Dashed blue line: the Sellke-Berger approach, using Appendix eqs. A8 and A7. Dotted blue line: the Goodman approach calculated using Appendix eqs, A9 and A7. Dotted red line is where points would lie if FPR was equal to the *p* value. This example is calculated for $n = 16$, normalised effect size = 1 and prior probability $P(H_1) = 0.5$. Log-log plot. Calculated with *Plot-FPR-vs-Pval+Sellke-Goodman.R* (supplementary material).

| Observed *p* value | Likelihood ratio: (H1/H0) | Prior for FPR=0.05 | Minimum false positive risk (mFPR) prior = 0.5 | | | False positive risk FPR with prior = 0.1 | | |
|---|---|---|---|---|---|---|---|---|
| | | | Colquhoun value | Sellke-Berger value | Goodman value | Colquhoun value | Sellke-Berger value | Goodman value |
| 0.05 | 2.8 | 0.87 | 0.27 | 0.29 | 0.227 | 0.76 | 0.79 | 0.725 |
| 0.025 | 6.1 | 0.76 | 0.14 | 0.20 | 0.140 | 0.60 | 0.69 | 0.593 |
| 0.01 | 15 | 0.55 | 0.061 | 0.11 | 0.068 | 0.37 | 0.53 | 0.395 |
| 0.005 | 29 | 0.40 | 0.034 | 0.067 | 0.037 | 0.24 | 0.39 | 0.259 |
| 0.001 | 100 | 0.16 | 0.0099 | 0.018 | 0.0088 | 0.082 | 0.14 | 0.074 |

Table 1.

Calculations done with web calculator http://fpr-calc.ucl.ac.uk/ , or with R scripts in supplementary materials. The method of calculation is described in the Appendix.

Column 1: The observed *p* value

Column 2: Likelihood ratios (likelihood for there being a real effect divided by likelihood of null hypothesis). Calculated for a well-powered (0.78) experiment, as in Appendix eq. A5 and Colquhoun (2017).

Column 3: The prior probability of there being a real effect, $P(H_1)$, that it would be necessary to postulate in order to achieve a false positive risk (FPR) of 5%. Calculated for a well-powered (0.78) experiment as in Appendix, eq. A7. (The reverse Bayes approach: Colquhoun, 2017).

Columns 4 - 6  The minimum false positive risk, i.e. the FPR that corresponds to a prior probability of $P(H_1) = 0.5$, calculated by three methods, Colquhoun value (calculated for a well-powered (0.78) experiment as in equations (A5) and (A6) and Colquhoun (2017). Sellke-Berger value: calculated as in Appendix eqs. (A8) and (A6). Goodman value, calculated as in Appendix eqs. (A9) and (A6).

Column 7 - 9. As columns 4 - 6, but for an implausible hypothesis with prior probability, $P(H_1) = 0.1$

The remaining columns of Table 1, show the false positive risks that correspond to each of the *p* values in column 1. They are calculated as described in the appendix, for prior probabilities, $P(H_1) = 0.5$ and for $P(H_1) = 0.1$.

It's gratifying that all three methods of calculating the FPR give results that are close enough to each other that much the same conclusions would be drawn in practice. All of them suggest that if you observe *p* = 0.05 in an unbiased, well-powered experiment, and you claim a discovery on that basis, you are likely to be wrong at least 20 -30% of the time, and over 70% of the time if the hypothesis is *a priori* implausible: ($P(H_1) = 0.1$).

## 8. Real life examples

*Example 1. The data used by Student (1908)*

| Drug A  | Drug B  |
| $y_A$   | $y_B$   |
|---------|---------|
| +0.7    | +1.9    |
| −1.6    | +0.8    |
| −0.2    | +1.1    |
| −1.2    | +0.1    |
| −0.1    | −0.1    |
| +3.4    | +4.4    |
| +3.7    | +5.5    |
| −0.8    | +1.6    |
|  0.0    | +4.6    |
| +2.0    | +3.4    |

Table 2.

Response in hours extra sleep (compared with controls) induced by (-)-hyoscyamine (A) and (-)-hyoscine (B). From Cushny and Peebles (1905).

The results in Table 2 are observations made by Cushny and Peebles (1905) on the sleep-inducing properties of (- )-hyoscyamine (drug A) and (-)-hyoscine (drug B). These data were used W. S. Gosset as an example to illustrate the use of his *t* test, in the paper in which the test was introduced (Student, 1908). The conventional *t* test for two independent samples is done as follows (more detail in Colquhoun, 1971, section 9.4).

>Number per sample: $n_A = 10$, $n_B = 10$
>
>Mean response: drug A: +0.75; drug B +2.33
>
>Standard deviations: drug A $s(y_A) = 1.78901$ ; drug B $s(y_B) = 2.002249$
>
>Effect size (difference between means): $E = 2.33 - 0.75 = 1.58$

We wish to test whether drug B is really better than A or whether the results could have plausibly arisen if the null hypothesis (true difference is zero) were true.

>The pooled estimate of the error within groups is

$$s(y) = \sqrt{\frac{(n_A - 1)s(y_A)^2 + (n_B - 1)s(y_B)^2}{(n_A - 1) + (n_B - 1)}} = 1.899$$

Normalised effect size = $E / s(y)$ = 0.83218

Standard deviation of effect size $s_E = \sqrt{\frac{s(y)^2}{n_A} + \frac{s(y)^2}{n_B}} = 0.84909$

Degrees of freedom: $(n_A - 1) + (n_B - 1) = 18$

Student's *t*: $t_{18} = E/s_E = 1.8608$

P = 0.07918

The *p* value exceeds 0.05 so the difference between means would conventionally be described as "non-significant". A very similar *p* value (0.08) is obtained from the same data by the randomisation test (Colquhoun, 1971, section 9.2, or, in more detail, in Colquhoun (2015), in which all 184,756 ways of selecting 10 observations from 20 are plotted.

This conventional *p* value result can now be compared with the false positive risk result. This can be done most conveniently by use of the web calculator (http://fpr-calc.ucl.ac.uk/ ). The inputs for the calculator are the *p* value (0.07918), the normalised effect size (0.83218) and the sample size (*n* = 10). If the prior probability of a real effect were 0.5 then the minimum false positive risk (mFPR) would be 0.28, a lot bigger than the *p* value (and it would be bigger still if the hypothesis were thought to be implausible. i.e. if its prior probability were less than 0.5). Another way to express the uncertainty is to note that in order to make the false positive risk 5% we would have to postulate a prior probability of there being a real effect, $P(H_1)$ = 0.88. Without strong prior evidence, the existence of a real effect is very doubtful. A third way to put the result is to say that the likelihood ratio is only 2.54 so the evidence of the experiment makes the existence of a real effect only 2.54 times as likely as it was before the experiment

The calculator shows also that the power of this study, calculated for *p* = 0.05 using the observed response size and standard deviation, is only 0.42.

*Example 2. A study of transcranial electromagnetic stimulation*

A study of transcranial electromagnetic stimulation (TMS), published in *Science*, concluded that TMS "improved associative memory performance", *p* = 0.043 (Wang et al. 2014).

The authors of this paper said only "The increase in performance for baseline to Post-Tx was greater for stimulation than for sham [T(15)= 2.21, P = 0.043]".  The mean raw scores (numbers of words correctly recalled to associated face cues) are given in their Table S2 in the supplementary material.  The calculations use the change in score between baseline and post-treatment. The effect size is the difference in this quantity between stimulation and sham. From Table S2, the effect size is 1.88 with standard deviation 1.70 ($n$ = 8).  The normalised effect size is therefore 1.88 / 1.70 = 1.10, and the *post hoc* power is 0.54.   Rather than saying only "*p* = 0.043", better ways to express the result of this experiment would be as follows (found from the [web calculator](#), radio buttons 3, 1 and 2, respectively). In all cases, the effect size and its confidence interval should be given in the text, not relegated to supplementary material, as follows.

The increase in memory performance was 1.88 ± 0.85 (stand deviation of the mean, *i.e.* standard error) with confidence interval 0.055 – 3.7 (extra words recalled on a baseline of about 10). Thus *p* = 0.043.

This should be supplemented as follows.

(a). The observation of *p* = 0.043  implies a minimum false positive risk (that for prior probability, $P(H_1)$ = 0.5), of at least 18%.  This is the probability that the results occurred by chance only, so the result is no more than suggestive.

or

(b). The increase in performance gave *p* = 0.043. In order to reduce the false positive risk to 0.05 it would be necessary to assume that we were almost certain (prior probability, $P(H_1)$ = 0.81)  that there was a real effect before the experiment was done.  There is no independent evidence for this assumption so the result is no more than suggestive.

or

(c). The increase in performance gave *p* = 0.043. In order to reduce the minimum false positive risk to 0.05 it would have been necessary to observe *p* = 0.0043, so the result is no more than suggestive.

Of these options, the first is probably the best because it is the shortest, and because it avoids the suggestion that FPR = 0.05 is being suggested as a new "bright line", in place of *p* = 0.05.

## 9. Conclusions

In my proposal, the terms "significant" and "non-significant" would not be used at all. This change has been suggested by many other people, but these suggestions have had little or no effect on practice.   A *p* value and confidence interval would still be stated but they would be supplemented by a single extra number that gives a better measure of the strength of the evidence than is provided by the *p* value.

(a). This number could be the minimum false positive risk, i.e. that calculated on the basis that the prior probability of a real effect is 0.5 (the largest value that it's reasonable to assume without hard evidence for a higher value: see Figure 3 in Colquhoun, 2017).  This has the advantage that it's easy to understand. It has the disadvantage that it's safe only for plausible hypotheses.  Nonetheless it would be a lot better than giving only the *p* value and confidence interval.

(b) Or the additional number could be the prior probability that you would need to believe in order to achieve a false positive risk of, say, 0.05.  This has the disadvantage that most users are unfamiliar with the idea of prior probabilities which are, in most cases, subjective probabilities.  It also runs the risk that FPR = 0.05 will become a new "threshold", in place of *p* = 0.05.

Examples of these improved ways to express the strength of the evidence are given above, at the end of section 8.

I appreciate that my proposal is only one of many ways to look at the problem of false positives. Other approaches can give different answers.  But it has the virtue of being simple to understand and simple to calculate.  Therefore it has a chance of being adopted widely

All the results in this paper have been derived in the context of *p* values found from *t* tests of the difference between means of two independent samples.  The question of the extent to which they apply to *p* values found in other ways remains to be investigated.  Since the logical flaw in the use of *p* values as a measure of the strength of evidence is the same however they are derived, it seems likely that the false positive risks will be similar in other situations.

Of course any attempt to reduce the number of false positives will necessarily increase the number of false negatives.  In practice, decisions must depend on the relative costs (in reputation, or in money, or both) that are incurred by wrongly claiming a real effect when there is none, and by failing to detect a real effect when there is one.

## APPENDIX

In this appendix we give details of three methods of calculating the likelihood ratio, and hence the FPR, for tests of a point null hypothesis against a simple alternative.

### A1. The *p-equals* approach as used by Colquhoun (2017)

The critical value of the *t* statistic is calculated from the observed *p* value as

$$t_{crit} = qt(1 - \frac{p}{2}, df, ncp = 0)$$

(A1)

where $qt()$ is the inverse cumulative distribution of Student's *t* statistic (the notation is as in R). The arguments are the observed *p* value, *p,* the number of degrees of freedom, *df*, and the non-centrality parameter, *ncp* (zero under the null hypothesis)

Numerical values given in this section all refer to the example shown in Figure A1, which is reproduced from Colquhoun (2017). In this case $p = 0.05$, with 30 degrees of freedom so $t_{crit} = 2.04$.

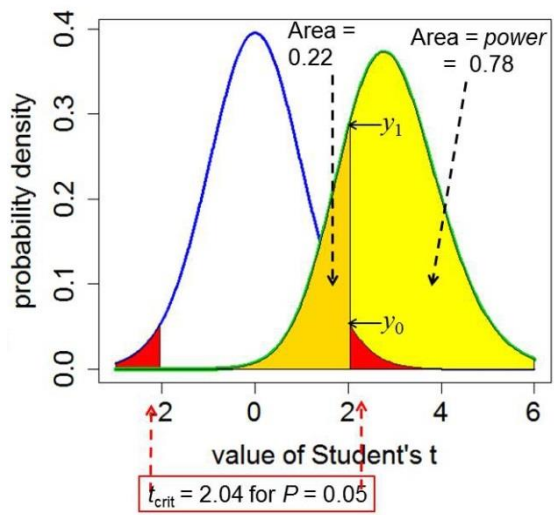

Figure A1.
Definitions for a null hypothesis significance test: reproduced from Colquhoun (2017). A Student's *t*-test is used to analyse the difference between the means of two groups of $n = 16$ observations. The *t* value, therefore, has $2(n − 1) = 30$ d.f. The blue line represents the distribution of Student's *t* under the null hypothesis ($H_0$): the true difference between means is zero. The green line shows the non-central distribution of Student's *t* under the alternative hypothesis ($H_1$): the true difference between means is 1 (1 s.d.). The critical value of *t* for 30 d.f. and $p = 0.05$ is 2.04, so, for a two-sided test, any value of *t* above 2.04, or below –2.04, would be deemed 'significant'. These values are represented by the red areas. When the alternative hypothesis is true (green line), the probability that the value of *t* is below the critical level (2.04) is 22% (gold shaded): these represent false negative results. Consequently, the area under the green curve above $t = 2.04$ (shaded yellow) is the probability that a 'significant' result will be found when there is in fact a real effect ($H_1$ is true): this is the *power* of the test, in this case 78%. The ordinates marked $y_0$ (= 0.526) and $y_1$ (= 0.290) are used to calculate likelihood ratios for the *p-equals* case.

Under the null hypothesis the probability density that corresponds to the observed *P* value is

$$y_0 = dt(tcrit, df, ncp = 0)$$

(A2)

where *dt*() is the probability density function of Student's *t* distribution with degrees of freedom *df* and non-centrality parameter = zero under the null hypothesis. This is the value marked $y_0$ in Figure A1, and in that example its value is 0.0526.

Under the alternative hypothesis, we need to use the non-central *t* distribution. The non-centrality parameter is

$$ncp = \frac{E}{s_E}$$

(A3)

where *E* is the difference between means (1 in the example) and $s_E$ is the standard deviation of the difference between means, 0.354, so *ncp* = 2.828. The probability density that corresponds with $t_{crit}$ = 2.04 is

$$y_1 = dt(tcrit, df, ncp = 2.828)$$

(A4)

In the example in Figure A1 this is $y_1$ = 0.290.

The likelihood of a hypothesis is defined as a number that is directly proportional to the probability of making the observations when the hypothesis is true. For the *p-equals* case we want the probability that the value of *t* is equal to that observed. This is proportional to the probability density that corresponds to the observed value. (More formally, the probability is the area of a narrow band centred on the observed value, but we can let the width of this band go to zero.) For a two-sided test, under the null hypothesis the probability occurs twice, once at *t* = – 2.04 and once at +2.04. Thus, under the *p-equals* interpretation, the likelihood ratio in favour of $H_1$, is

$$L_{10} = \frac{L(H_1)}{L(H_0)} = \frac{Prob(data \mid H_1)}{Prob(data \mid H_0)} = \frac{y_1}{2y_0}$$

(A5)

In the example in Figure A1 the likelihood ratio is 0.290 / (2 x 0.0526) = 2.76. The alternative hypothesis is 2.76 times more likely than the null hypothesis.

The likelihood ratio can be expressed as a false positive risk via eq 5 in the text

$$FPR = \frac{1}{1 + L_{10}\frac{P(H_1)}{1 - P(H_1)}}$$

(A6)

The reverse Bayes approach expresses the result as the prior probability, $P(H_1)$, that would be needed to get a specified FPR. This can be found by solving eq. A6 for $P(H_1)$

$$P(H_1) = \frac{(1 - FPR)}{(1 - FPR) + L_{10}FPR}$$

(A7)

## A2. The Sellke – Berger approach

Sellke *et al*, (2001) proposed to solve the problem of the unknown prior distribution by looking for an upper bound for the likelihood ratio for $H_1$ relative to $H_0$. Expressed as a function of the observed *p* value, they suggest

$$L_{10} = \frac{1}{-ep \log(p)}$$

(A8)

(this holds for $p < 1/e$, where $e = 2.71828 \ldots$). This is the largest odds in favour of rejection of the null hypothesis, $H_0$, that can be generated by any prior distribution, whatever its shape. It is the choice that most favours the rejection of the null hypothesis. It turns out that nevertheless, it rejects the null hypothesis much less often than the *p* value. As before, the likelihood ratio from eq. A8 can be converted to a false positive risk via eq. A6, or to the reverse Bayes approach with eq.A7.

For the case where the prior probability of having a real effect is $P(H_1) = P(H_0) = 0.5$, this can be interpreted as the minimum false discovery rate (Sellke *et al.*, 2001). It gives the *minimum* probability that, when a test is 'significant', the null hypothesis is true: i.e. it is an estimate of the minimum false positive risk.

The FPR can be calculated by substituting the likelihood ratio from from eq.A8 into eq. A6, and the reverse Bayesian calculation of $P(H_1)$ can be found by substituting this FPR into eq. A7.

## A3. The Goodman approach

For large samples, such that the *t* distribution is well approximated by the normal distribution, Goodman (1999b) gives the maximum likelihood ratio in favour of $H_1$ as

$$L_{10} = \frac{1}{2\exp(-z^2/2)}$$

(A9)

(This is the likelihood ratio calculated at the sample mean, the maximum likelihood estimate of the true effect size.)

In equation A9, *z* is the observed standard normal deviate, which is related to the observed *p* value, according to

$$z = qnorm(1 - \frac{p}{2}, 0, 1)$$

Here *qnorm* is the inverse cumulative probability distribution function for a standard normal distribution (the notation is as in R). The factor of 2 on the bottom line of eq.

A9 is needed since we are talking about a 2-sided test: see Held & Ott, 2018, equation 13).

The FPR can be calculated by substituting $L_{10}$ from eq. A9 into eq. A6. The reverse Bayesian calculation of $P(H_1)$ can be found by substituting this FPR into eq. A7.

### Acknowledgements

I am especially grateful to the following people.

Dr Leonhard Held Epidemiology, Biostatistics and Prevention Institute, University of Zurich, CH-8001 Zurich, for helpful discussions/

Dr R.A.J. Matthews for helpful discussions about his paper (Matthews 2018).

Prof Stephen Senn (Competence Center in Methodology and Statistics, CRP-Santé, Luxembourg)

Prof D. Spiegelhalter, for reading an earlier version of Colquhoun (2017) and suggesting the term "false positive risk".

### Supplementary material

The R scripts, and the output text files that they generate, can be downloaded as a single compressed zip file, at
http://onemol.org.uk/pvals-2018/Supp-files-p-vals-2018.zip